\documentclass[%
reprint,
superscriptaddress,
amsmath,amssymb,
aps,
pra,
showpacs,
showkeys,
onecolumn
]{revtex4-2}
\usepackage{graphicx}
\usepackage{dcolumn}
\usepackage{bm}
\usepackage{braket}
\usepackage{color}
\usepackage{hyperref}
\usepackage{listings}
\usepackage{caption}
\usepackage{subcaption}

\begin{document}

\title{Collapse and revivals for the binomial field distribution}

\author{S.I. Pavlik}
\email{sipavlik@yahoo.com}
\affiliation{Independent Researcher, Zaporizhzhia 69104, Ukraine}

\date{\today}

\begin{abstract}
The exact representation of the atomic inversion in the Jaynes-Cummings model as an integral over the Hankel contour is used. For a field in a binomial state, the integral is evaluated using the saddle point method. Simple approximate analytical expressions for collapse and revivals are obtained.

\end{abstract}

\pacs{42.50.-p, 42.50.Ar, 42.50.Dv}

\keywords{Jaynes-Cummings model, atomic inversion, binomial distribution, collapse and revival, saddle-point method}
\maketitle


The Jaynes-Cummings model consists of the two-level system with level separation $\omega_0$ interacting with a single-mode quantized electromagnetic field with a frequency $\omega$. In the rotating wave approximation, this model is described by the Jaynes-Cummings Hamiltonian \cite{jaynes1963comparison,Jay} for which the interaction may be written as 
\begin{equation}
	H_{I}=\lambda(\sigma_{+}a+\sigma_{-}a^{+}),\label {eq:E1}
\end{equation}
where $a$ and $a^{+}$ are the annihilation and creation operators of the photon field, $\lambda$  is the coupling constant. We use the Pauli matrices $\sigma_{i}$   ($i=1,2,3$ ) to describe the two-level atom and the notation $\sigma_{\pm}=1/2(\sigma_{1} \pm \sigma_{2})$. For simplicity we have assumed the coupling $\lambda$ to be real. Various aspects of the Jaynes-Cummings model are reviewed in numerous reviews, see for example \cite{ShoKni, larson2021jaynes}.
 
The Jaynes-Cummings Hamiltonian (\ref{eq:E1}) is easily diagonalizable \cite{Jay}, since it connects only a pair of basis states $| n \rangle |1\rangle\rightleftarrows| n-1 \rangle |2\rangle$, where $|1\rangle$  and $|2\rangle$  are the ground and exited state, and $|n\rangle$ is an eigenstate of the photon-number operator $a^{+}a$. If we assume that the atom is initially in the ground state $|1\rangle$, and the field is in a state $\sum_{n=0}^N\gamma_{n}|n\rangle$, where $W_{n}=|\gamma_{n}|^2$ determines the distribution of photons, the atomic inversion in the resonant case ($\omega_0=\omega$) is given by
\begin{equation}
	\langle\sigma_{3}(t)\rangle=-\sum_{n=0}^{N} W_{n}\cos(2\sqrt{n}\lambda t).\label {eq:E2}
\end{equation}
In the follows, we will use the replacing $\lambda t\rightarrow t$.

If the field is initially prepared in a coherent state, then the initial value of the atomic inversion quickly decays, it has been called Cummings collapse \cite{Cum}, but it reappears many times at a later time \cite{EbNaSa, NarSaEb}. The envelope of the Rabi oscillations periodically revives and collapses. These phenomena occurs in the real world and has been observed in experiments \cite{PhysRevLett.58.353, PhysRevLett.76.1800, PhysRevLett.76.1796}.

For the binomial state of the field \cite{miller1967anti, stoler1985binomial},
\begin{equation}
W_{n}=\frac{N!}{(N-n)!n!}p^{n}q^{N-n},\label {eq:E3}
\end{equation}
where $0\leq p\leq1$ and $q=1-p$, the phenomenon of collapse and revival was studied in detail in \cite{joshi1987effects, joshi1989effects1}. Our aim in this paper is to investigate the phenomenon of collapse and revival by considering the field to be in the binomial state, according to the method proposed in \cite{pavlik2023inside}.
 \paragraph*{Integral representation of the atomic inversion.}

Since the summation index under the square root, it is desirable to use such a representation of the trigonometric function so that the replacement occurs $\sqrt{n}\rightarrow n$. Taking into account the variety of integral representations of Bessel functions, we choose the equality $\cos{x}=\sqrt{\pi x/2}J_{-1/2}(x)$ \cite{WhiWa}, where $J_{-1/2}(x)$ is the Bessel functions of half-integer order. This is motivated by the fact that the Bessel functions have a suitable representation as a contour integral in which the argument is squared.
Using the integral representation of the Bessel function by inverting the order of integration and summation, the Jaynes-Cummings sum (\ref{eq:E2}) is rewritten as \cite{pavlik2023inside}
\begin{align}
	&\langle\sigma_{3}(t)\rangle=-\frac{1}{2\sqrt{\pi}i}
	\int_{-\infty}^{(0^+)}\frac{e^{z}}{\sqrt{z}}\sum_{n=0}^N W_{n} e^{-nt^{2}/z}dz,\label {eq:E4}
\end{align}
where $\mid\arg z \mid<\pi$; the cut is along the negative real semi-axis. The path of integration is the Hankel contour coming from $-\infty$, turning upwards around $0$, and heading back towards $-\infty$ \cite{Cop}. The notation $(0^+)$ just means that the path goes round the origin in the positive sense. 

In the representation (\ref{eq:E4}) the sum is greatly simplified, since the generating function of the distribution of photons obtained under the integral sign is easily calculated for most probability distributions.
In addition, the convergence of the integral in Eq.~(\ref{eq:E4}) at the ends of the path is exponential. 
\paragraph*{Binomial photon distribution.}
If the field is initially in a binomial state (\ref{eq:E3}), calculating the sum in Eq.~(\ref{eq:E4}), the atomic inversion is given by
\begin{equation}
\langle\sigma_{3}(t)\rangle=-\frac{N^{1/2}\tau^{1/2}}{2\sqrt{\pi}i}
\int_{-\infty}^{(0^+)}\frac{1}{\sqrt{z}}e^{N\Phi(z)}dz,\label {eq:E5}
\end{equation}
with
\begin{equation}
	\Phi(z)=\tau z+\ln{(pe^{-1/z}+q)},\label {eq:E6}
\end{equation}
where $\tau=t^{2}/N$ and $z\rightarrow zt^{2}$. Note that the origin is both a branch point and an essential singularity. 

For large $N$, we use the saddle point approximation to the integral in Eq.~(\ref{eq:E5}). Numerous examples of asymptotic analysis of special functions based on the contour integral representation can be found in the book \cite{Olver}.

Saddle points are found from the equation
$\Phi^{'}(z_{0})=0$, where the prime denotes a derivative with respect to $z$. Solutions to this equation give paths, $z_{0}(\tau)$, in the complex plane. The contribution to the atomic inversion is 
\begin{equation}
	\langle\sigma_{3}(t)\rangle\cong i\sqrt{\frac{\tau}{2|\Phi^{''}(z_{0})|z_{0}}} e^{N \Phi(z_{0})+i\theta}, \label {eq:E7}
\end{equation}
where the angle is determined from the condition $2\theta+arg\Phi^{''}(z_{0})=\pi+2\pi m$ ($m$ is an integer). If we introduce a new variable $W=-1/z_{0}$ and the notation $\alpha=q/p$, then
\begin{equation}
	\phi(W)=\Phi\left(-\frac{1}{W}\right)=-\frac{\tau}{W}+ \ln{p}+\ln{(e^{W}+\alpha)} \label {eq:E8}
\end{equation}
and
\begin{equation}
	\Phi^{''}=-2W\tau f,\label {eq:E9}
\end{equation}
with
\begin{equation}
	f=1+\frac{1}{2}\left(W+\frac{\tau}{W}\right). ,\label {eq:E10}
\end{equation}
Here $W$ is a solution of the equation
\begin{equation}
	\tau=-\frac{W^{2}e^{W}}{e^{W}+\alpha},\label {eq:E11}
\end{equation}
which is obtained from $\Phi^{'}=0$. Using Eq.~(\ref{eq:E8}) and Eq.~(\ref{eq:E9}) in Eq.~(\ref{eq:E7}), we get
\begin{equation}
	\langle\sigma_{3}(t)\rangle\cong\frac{i}{2}\sqrt{\frac{-W}{|Wf|}} e^{N\phi(W)+i\theta} +\text c.c., \label {eq:E12}
\end{equation} 
where '$\text c.c.$' denotes the complex conjugate, this follows from the fact that the atomic inversion is a real function.
The angle $\theta$ is defined as $\theta=\pi/2 -arg(-W)/2-argf/2$, where we have used Eq.~(\ref{eq:E9}).
Thus, the basic expression for the atomic inversion associated with a certain saddle point trajectory $W(\tau)$ is given by
\begin{equation}
	\langle\sigma_{3}(t)\rangle\cong -\frac{1}{\sqrt{|f|}} e^{N \Re\phi(W)} cos\left(N\Im\phi(W)-\frac{1}{2}argf\right), \label {eq:E13}
\end{equation}
where the functions $\Re\phi(W)$ and $\Im\phi(W)$ are the real and imaginary parts of $\phi(W)$. This is a universal expression that describes both collapse and revival. Now we need to find solutions of Eq.~(\ref{eq:E11}) at different times and use it in Eq.~(\ref{eq:E13}).
\paragraph*{Collapse.}
Consider the trajectory of the saddle point $W(\tau)$ for small $\tau$. Since the right side of the Eq.~(\ref{eq:E11}) is regular in the neighborhood $W=0$, we can use the Lagrange's formula for the reversion of a power series \cite{Cop}, that is,  
\begin{equation}
	W(u)=\sum_{n=1}^{\infty}c_{n}(i\tau^{1/2})^n,\label {eq:E14}
\end{equation}
where 
\begin{equation}
	c_{n}=\frac{1}{n!}\frac{\mathrm{d}^{n-1}}{\mathrm{d}W^{n-1}}\left(1+\alpha e^{-W}\right)^{n/2}\bigg\vert_{W=0}.
\end{equation}
For $\tau\ll1$, you can take only the first two terms in Eq.~(\ref{eq:E14}), $W(\tau)\approx i\tau^{1/2}(1+\alpha)^{1/2}+\alpha\tau/2$. Substituting this expression in Eq.~(\ref{eq:E8}) we immediately arrive at
\begin{align}
	\phi\approx-\tau q/2+2i\tau^{1/2}p^{1/2}.\label {eq:E15}
\end{align}
Since here $\Im W\ll1$, then $arg f\approx 0$, Eq.~(\ref{eq:E13}) rewritten as
\begin{equation}
	\langle\sigma_{3}(t)\rangle\approx- e^{-t^2 q/2} cos\left(2N^{1/2}p^{1/2}t\right). \label {eq:E16}
\end{equation}
This is a well-known expression for the Cummings collapse for the binomial field distribution. Note that $q=0$ in Eq.~(\ref{eq:E16}) corresponds to a pure state which results in simple Rabi oscillations of frequency $2N^{1/2}$,  as follows from Eq.~(\ref{eq:E2}) and Eq.~(\ref{eq:E3}).
\paragraph*{Revivals.}
Now consider the contribution to the atomic inversion (\ref{eq:E16}) from saddle points on other branches of the function $W$. It is easy to see that $\Re\phi$ takes its maximum value when $\tau_{n}=4\pi^2 p n^2$, $W(\tau_{n})=W_{n} =i 2\pi n$, where $n$ is integer (see Eq.~(\ref{eq:E11})). Note that Eq.~(\ref{eq:E11}) also has a solution for half-integer $n$ and $\alpha<1$, but, unfortunately, it is forbidden by the logarithm in Eq.~(\ref{eq:E6}). On a $z$-plane, a point $W_{n}$ ($n<0$) corresponds to a point that is on the negative imaginary axis. The steepest descent path intersects the axis at $\theta<\pi/2$. Strictly speaking, we must take $W_{n}$ in Eq.~(\ref{eq:E11}) for $n< 0$ (in this case $\theta<\pi/2$), but since a pair of complex conjugate saddle points contribute to the final result, we will use $W$ for any positive values $n$.

The Taylor series of a function in Eq.~(\ref{eq:E8}) about a point $\tau=\tau_{n}$ is given by
\begin{equation}
	\phi(\tau,W(\tau))\approx\phi(\tau_{n},W_{n})+ \frac{\mathrm{d}\phi}{\mathrm{d}\tau}\bigg\vert_{\tau_{n}}(\tau-\tau_n)+\frac{1}{2}\frac{\mathrm{d}^2\phi}{\mathrm{d}\tau^2}\bigg\vert_{\tau_{n}}(\tau-\tau_n)^2+\dots \label {eq:E17}
\end{equation} 
where the total derivatives are taken along the curves defined by the equation $\phi_{W}=0$, that is, 
\begin{align}
	\frac{\mathrm{d}\phi}{\mathrm{d}\tau}=\phi_{\tau},\;\;\frac{\mathrm{d}^2\phi}{\mathrm{d}\tau^2}=\phi_{\tau\tau}+\phi_{W\tau}W_{\tau}=-\frac{\phi_{W\tau}^2}{\phi_{WW}},\nonumber
\end{align}
where $W_{\tau}=-\phi_{W\tau}/\phi_{WW}$; we used the rule of implicit differentiation. For $\tau$ near  $\tau_n$ in Eq.~(\ref{eq:E17}), one can replace $(\tau-\tau_n)^2\approx 4\tau_n(\tau^{1/2}-\tau_n^{1/2})^2$. The derivatives of $\phi$ at $\tau_n$ are
\begin{equation}
	\phi_{\tau}\vert_{\tau_{n}}=-1/W_n,\;\;\phi_{W\tau}\vert_{\tau_{n}}=1/W^{2}_n,\;\;\phi_{WW}\vert_{\tau_{n}}=-\frac{2\tau_n}{W^{3}_n}+qp.\nonumber
\end{equation}
Collecting the derivatives of $\phi$ in Eq.~(\ref{eq:E17}), substituting into Eq.~(\ref{eq:E13}), we obtain
\begin{equation}
	\langle\sigma_{3}(t)\rangle\approx-\frac{1}{(1+\pi^2 q^2 n^2)^{1/4}} \exp{\left[-\frac{q(t-t_n)^2}{2(1+\pi^2 q^2 n^2)}\right]}\cos\left[\frac{t^2}{2\pi n}-\frac{(t-t_n)^2}{2\pi n(1+\pi^2 q^2 n^2)}-\frac{1}{2}arg(1+i\pi n q)\right],\label {eq:E18}
\end{equation}
where $t_n = nT$, $T=2\pi N^{1/2} p^{1/2}$ is a period of revivals. 
Recall that Eq.~(\ref{eq:E10}) implies $f(\tau_n,W_n)=1+i\pi n$. For not too small $q$ we have $arg(1+i\pi n q)\approx\pi/2$.
\begin{figure}
	\centering
	\begin{subfigure}[b]{0.45\textwidth}
		\includegraphics[width=\textwidth]{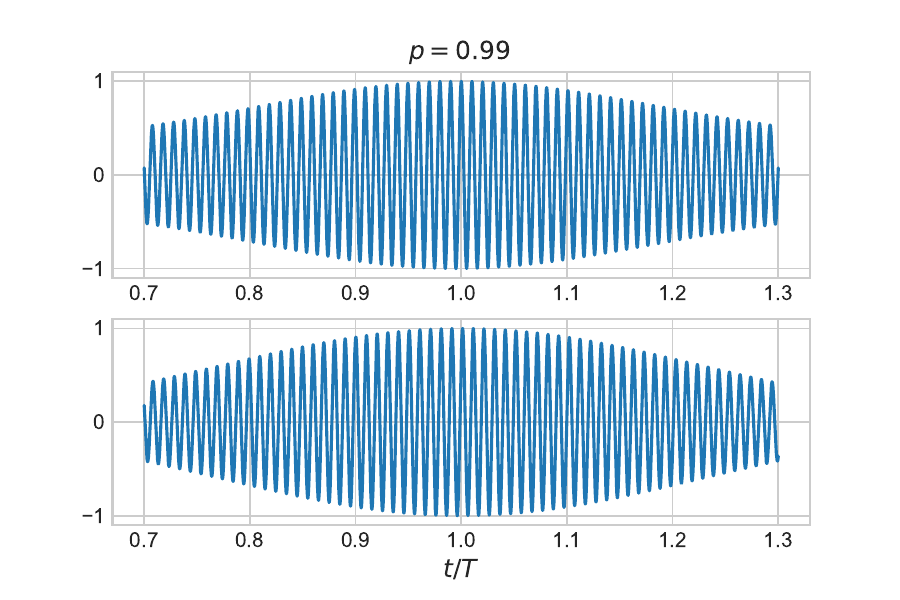}
		\caption{}
		\label{fig:a}
	\end{subfigure}
	\begin{subfigure}[b]{0.45\textwidth}
		\includegraphics[width=\textwidth]{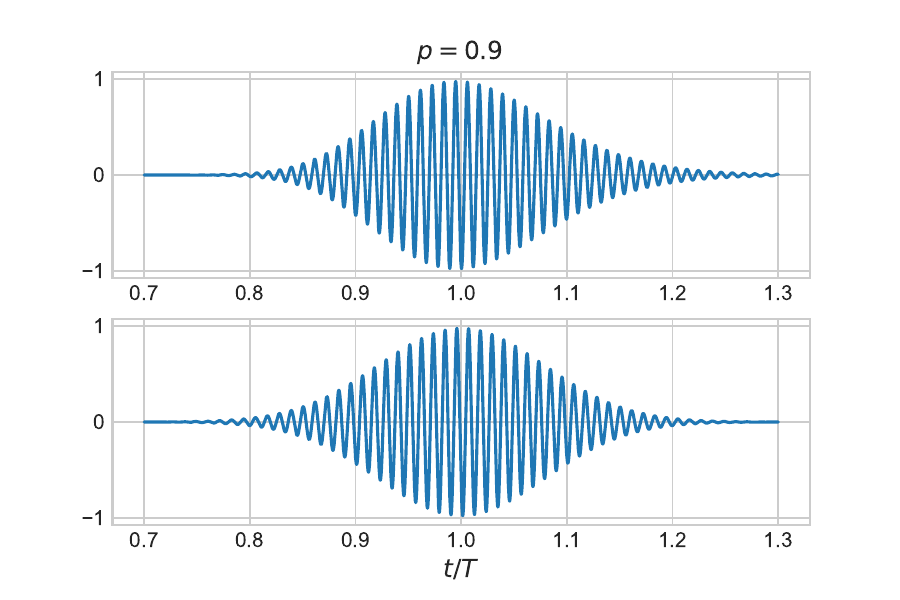}
		\caption{}
		\label{}
	\end{subfigure}
	\begin{subfigure}[b]{0.45\textwidth}
		\includegraphics[width=\textwidth]{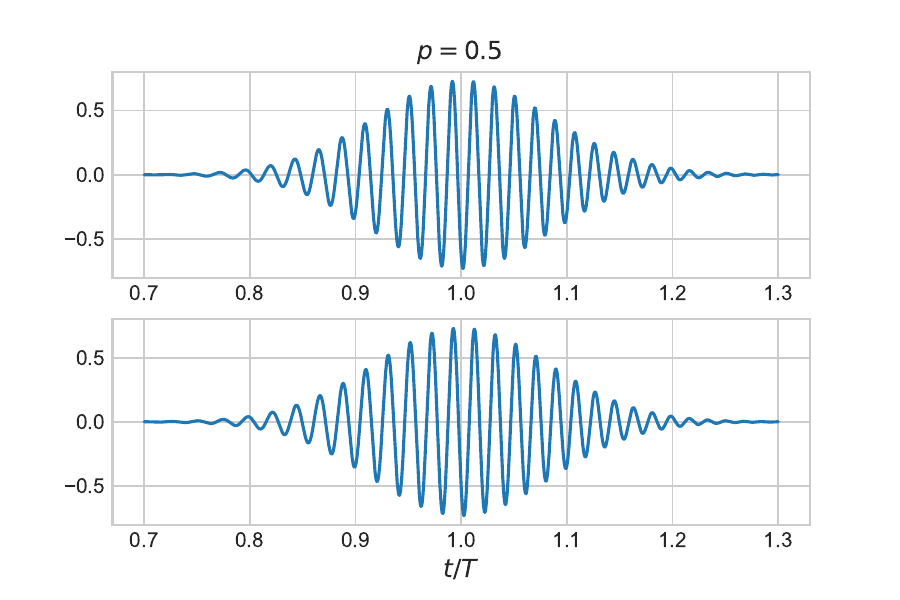}
		\caption{}
		\label{}
	\end{subfigure}
	\begin{subfigure}[b]{0.45\textwidth}
		\includegraphics[width=\textwidth]{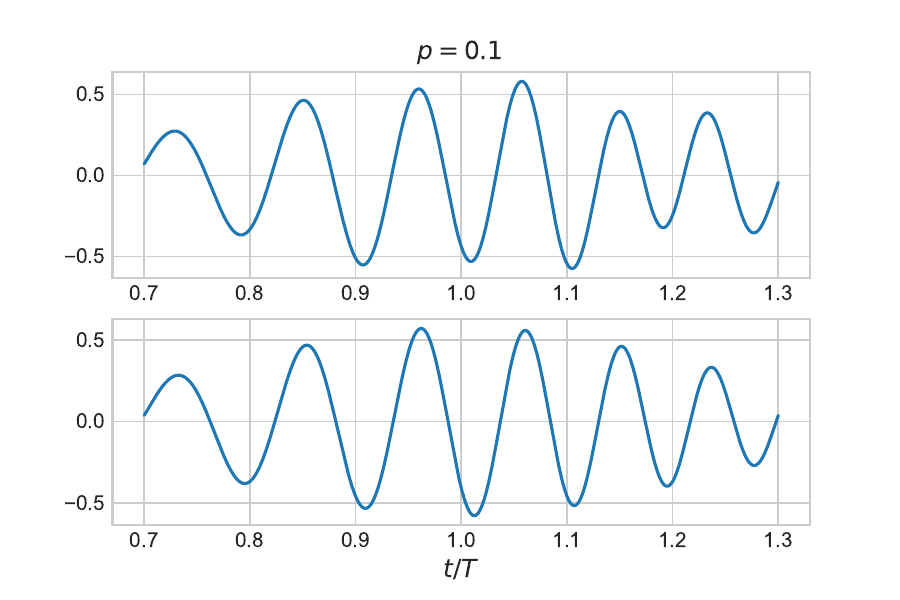}
		\caption{}
		\label{}
	\end{subfigure}	
	\caption{First quantum revival for a binomial state as a function of the scaled time $t/T$ ($T=2\pi N^{1/2} p^{1/2}$) for $N=50$. The atomic inversion $\langle\sigma_{3}\rangle$ is plotted on the vertical axes. Curves a, b, c and d are for different $p=1-q$. In each box, the top graph is from the exact sum Eq.~(\ref{eq:E2}) and the bottom graph is from the approximate result Eq.~(\ref{eq:E18}).}
	\label{fig}
\end{figure}

Figure~\ref{fig} shows the atomic inversion $\langle\sigma_{3}\rangle$ as a function of scaled time $t/T$. In order to compare the exact and approximate solution, we consider the first quantum revival ($n=1$). As can be seen, the approximate analytical result based on Eq.~(\ref{eq:E18}) for $n=1$ (the bottom graph in each boxes) is in good agreement with the exact expression Eq.~(\ref{eq:E2}) (top graphs). Note that we took the last expression under the cosine equal to $\pi/4$. As the value decreases, this angle tends to zero, which leads to a phase shift in Figure~\ref{fig:a}.
\paragraph*{Discussions and summary.}
In the resonance case, using the integral representation for the Bessel function of half-integer order, the Jaynes-Cummings sum was written as a contour integral of the generating function for the photon distribution  \cite{pavlik2023inside}. For a field in a binomial state, an exact integral representation of the Jaynes-Cummings sum is obtained. For large $N$, the saddle method was used to obtain an approximate expression for the integral. Simple approximate analytical expressions for collapse and revivals are obtained. Since the approximate solution looks more complicated and cumbersome, in many cases it is advisable to use the exact expression. However, on the way to an approximate solution, we discovered the essence of this strange world. The magic of collapse and revivals is hidden in the solution $W(\tau)$ of Eq.~(\ref{eq:E11}). For $q<p$, a new solution of the equation appears for which $W=i\pi(2n+1)$. There is hope for a phase transition. However, such a solution is forbidden by the logarithm in Eq.~(\ref{eq:E6}). Is it possible to find a photon distribution such that new solutions of the saddle point equation arise when the parameters change?
\begin{acknowledgments}
It is a pleasure to thank Moorad Alexanian. I’d like to express gratitude to my family.
\end{acknowledgments}

\bibliography{jcb}

\end{document}